\documentclass[twocolumn,showpacs,preprintnumbers,amsmath,amssymb,superscriptaddress]{revtex4}
\usepackage{graphicx}
\usepackage{dcolumn}
\usepackage{bm}
\usepackage{natbib}

\def\ltsima{$\; \buildrel < \over \sim \;$}
\def\simlt{\lower.5ex\hbox{\ltsima}}
\def\gtsima{$\; \buildrel > \over \sim \;$}
\def\simgt{\lower.5ex\hbox{\gtsima}}

\begin{document}
\title{Illuminating Dark Energy with Cosmic Shear}

\author{Fergus Simpson}
\email{frgs@ast.cam.ac.uk} \affiliation{Institute of Astronomy,
University of Cambridge, Madingley Road, Cambridge CB3 0HA}

\author{Sarah Bridle}
 \affiliation{Department of Physics and Astronomy, University College London, Gower Street,
London WC1E 6BT}


\date{\today}
\newcommand{\ud}{\mathrm{d}}

\begin{abstract}
  One of the principal goals of modern cosmology is to constrain the
  properties of dark energy. At present, numerous surveys are aiming
  to determine the equation of state, which is assumed constant due to
  our poor understanding of its behaviour (and since higher order
  parameterisations lead to unwieldy errors). This raises the question
  - how does our ``best-fit" equation of state relate to the true
  function, which may vary with redshift. Saini et al.  (2003)
  \cite{2003MNRAS.343..533D} have demonstrated that the value of $w$
  attained by a supernovae study is well described by a weighted
  integral over the true function. Adopting a similar approach, we
  calculate the corresponding ``weight function" for a fiducial cosmic
  shear survey. We disentangle contributions from the angular diameter
  distance and the growth factor, finding that they partially cancel
  each other out. Consequentially, the sensitivity to $w$ at high
  redshift is enhanced beyond that of supernovae studies, and is
  comparable, if not higher, than lensing of the CMB. This illustrates
  the complementary nature of the different techniques.  Furthermore,
  we find that results which would na\"{\i}vely be interpreted as
  inconsistent could arise from a time-dependent equation of state.

\end{abstract}

\maketitle

\section{\label{sec:level1}Introduction}

Following the rapid accumulation of experimental evidence
\cite{1999ApJ...517..565P, 2004MNRAS.tmp..258A, 2004PhRvD..69j3501T,
  2003ApJS..148..175S}, it has become generally accepted that the
universe is presently experiencing a period of accelerated expansion.
In the context of general relativity, this implies that current
cosmological dynamics are dominated by a component with negative
pressure. By revealing the behaviour of this dark energy, not only
might we foresee the ultimate fate of the universe, but this may be
the first step into a new field of physics.

The equation of state for dark energy, defined as the pressure to
density ratio $w=p/\rho$, provides us with an observable quantity
which may differentiate between the leading candidates. Any deviation
from $w=-1$ (the value indicative of the cosmological constant)
imparts two major influences in cosmology.  For a fixed dark energy
density, a more negative $w$ is associated with stronger pressure,
which in turn enhances the expansion of the universe.

The effect of this can be seen in cosmography (measurements of the
global properties of the universe), such as luminosity distance (in
the case of supernovae), and comoving angular diameter distance (in
the case of weak lensing).

Secondly, the dark energy density evolution is governed by $w$. A more
negative $w$ leads to the mass fraction of dark energy $\Omega_{\rm
  DE}$ decaying more rapidly with redshift. This alters both the
expansion and structure formation histories.

The results of dark energy studies may be concisely expressed as
an array of estimated values for the (constant) equation of state.
It is imperative that we are alert to the various interpretations
which may be drawn from this, bearing in mind the possibility that
$w$ is time-dependent. In this study we will concentrate on the
redshift sensitivity with which cosmic shear probes dark energy,
and the extent to which the data complements that of supernovae
surveys.

Cosmic shear is perhaps the most promising technique for quantifying
dark energy, largely due to the simplicity of the underlying physics.

Apparent alignments of distant galaxies arise from weak gravitational
lensing by intervening structure
\cite{2000A&A...358...30V,2000MNRAS.318..625B,2000Natur.405..143W,2003AJ....125.1014J,astro-ph/0003338}.
The strengths of these correlations are dictated by two cosmological
features: the matter power spectrum, and the lens geometry. In turn,
both of these depend on $w$. Here we study their relative
contributions to the shear signal, and show how this influences the
range of $w(z)$ to which cosmic shear is most sensitive.

The structure of this article is as follows. In \S II the theory
of cosmic shear is briefly reviewed, along with the approach we
will use to evaluate the weight function.

We present the weight function for a fiducial cosmic shear survey in
\S III, separating its contributions from the geometry and the growth
of structure. This is extended in \S IV to tomography, while \S V uses
an alternative method to investigate whether cosmic shear measures $w$
via its influence on structure growth or the expansion rate. Lensing
of the cosmic microwave background and cosmic shear surveys by the
Large Synoptic Survey Telescope (LSST) and Canada-France-Hawaii
Telescope Legacy Survey (CFHTLS) are considered in \S VI. Finally, we
compare the information which can be extracted from lensing and
supernova surveys in \S VII.

\section{The Sensitivity of Cosmic Shear}

Here we relate the underlying equation of state $w(z)$ to the constant
value $w$ which provides the best fit to cosmic shear data. The survey
parameters used are that of the proposed \emph{Supernovae/Acceleration
  Probe} (SNAP), although as we shall see in \S VI our results are
robust to the exact specifications, with the exception of the survey
depth.

\subsection{Methodology}

\subsubsection{The Matter Power Spectrum}

We adopt the following $\Lambda$CDM cosmology as our fiducial
model: $w=-1$, $\Omega_m=0.3$, spectral index $n=1$, baryon
fraction $\Omega_b=0.047$, $\sigma_8=0.88$, and Hubble parameter
$h=0.7$.

To produce the matter power spectrum $P(k,z)$ we use the shape
parameter of Sugiyama $[\Gamma = \Omega_m h
\exp[-\Omega_b(1+\sqrt{2h}/\Omega_m)]]$. Linear growth of structure is
evaluated in accordance with standard theory:

\begin{equation}
\ddot{\delta}_m  +
\frac{3}{2a}\big[1-w(a)(1-\Omega_m(a))\big]\dot{\delta}_m -
\frac{3}{2a^2}\Omega_m(a) \delta_m = 0
.
\end{equation}

For this study the large-scale clustering of dark energy is considered
to be of little significance \cite{1999ApJ...521L...1M}. We then apply
the fitting formula from Peacock \& Dodds \cite{1996MNRAS.280L..19P},
whilst acknowledging that a more sophisticated approach would be
preferred when modelling quintessence cosmologies.

\subsubsection{Weak Lensing}

The shear power spectrum is given by

\begin{equation} \label{eq:shear}
C_\ell = \frac{\displaystyle 9}{16}\Big(\frac{\displaystyle
H_0}{\displaystyle c}\Big)^4 \Omega_m^2 \int^\chi_0
\Big[\frac{\displaystyle g(\chi)}{\displaystyle
ar(\chi)}\Big]^2P(\frac{\ell}{r},\chi) \ud \chi
\end{equation}

\noindent where $\chi$ is the coordinate distance, and $r(\chi)$
the comoving angular diameter distance.  We assume a flat universe
throughout, and thus $r(\chi)=\chi$.  $P(\ell/r,\chi)$ is the matter
power spectrum for wavenumber $\ell/r$ at a redshift corresponding to
a coordinate distance $\chi$.  The lensing efficiency term $g(\chi)$
favours the density perturbations which are approximately equidistant
from the source and observer:
\begin{equation}
g(\chi)=2 \int_{\chi}^{\chi_h}
n(\chi')\frac{r(\chi)r(\chi'-\chi)}{r(\chi')} \ud\chi' ,
\label{eq:lenseff}
\end{equation}
where $n(\chi)$ is the galaxy probability density function, with
respect to comoving distance (see Bartelmann \& Schneider
\cite{1999A&A...345...17B} and Mellier \cite{1999ARA&A..37..127M} for
reviews).

Multipole errors are, in accordance with Kaiser
\cite{1998ApJ...498...26K}, given by
\begin{equation}
\sigma(\ell)= \sqrt{\frac{\displaystyle 2}{\displaystyle(2\ell
+1)f_{sky}}}(C_\ell+\frac{\displaystyle\sigma^2_\gamma}{\displaystyle
2n_g}) ,
\end{equation}
where $f_{sky}$ is the area expressed as a fraction of the whole sky,
$n_g$ is the number of galaxies per steradian and $\sigma_{\gamma}$ is
the rms intrinsic (pre-lensing) ellipticity of the galaxies in the
sample.

For the purposes of
our calculations we followed the fitting function and survey parameters from Refregier
et al. \cite{2004AJ....127.3102R}
\begin{equation} \label{eq:nofz}
n(z) \propto z^\alpha e^{-(z/z_0)^\beta} .
\end{equation}
For simulating the SNAP wide survey the following figures were used:
redshift parameter $z_0=1.13$, $\alpha=2$, $\beta=2$, $f_{sky}$ is the
equivalent of $1000$ square degrees, $n_g=100$ ${\rm arcmin}^{-2}$ and
$\sigma_\gamma=0.31$.

\label{sec:SNAPpars}

In light of an investigation by Cooray \& Hu
\cite{2001ApJ...554...56C}, we can afford to ignore the correlation of
errors for different $\ell$'s. Furthermore, we enforce an upper limit
of $\ell<10^4$, above which the baryons are considered to introduce a
significant uncertainty in the power spectrum estimation
\cite{2004APh....22..211W,2004ApJ...616L..75Z}. At the end of
\S~\ref{sec:weightfn} we investigate the effect of adopting a more
conservative upper limit.

\subsection{Response to the Equation of State}

The functional derivative describes the manner in which the shear
signal reacts to changes in the value of $w$ at various redshifts
\begin{equation}
\frac{\partial C_\ell(w^{\rm fid})}{\partial w(z)}\Bigg
\arrowvert_{z'} = \lim_{\epsilon \rightarrow 0}
\frac{C_\ell(w^{\rm fid}+\epsilon \delta(z-z'))
-C_\ell(w^{\rm fid})}{\epsilon}.
\end{equation}

Since $w$ only imparts a small change in the observed shear, we can
approximate the deviation from a fiducial model as
\begin{equation}
C_\ell \simeq C_\ell (w^{\rm fid}) + \int{\frac{\partial
C_\ell(w^{\rm fid})}{\partial w(z)}\Bigg\arrowvert_{z'}
(w(z')-w^{\rm fid}(z')) \ud z'}.
\end{equation}

Before proceeding further we introduce the notation $\delta w(z') =
w(z')-w^{\rm fid}(z')$, $\Delta w = w^{\rm fit}-w^{\rm fid}$, and
$K(\ell,z')=\partial C_\ell(w^{\rm fid})/\partial w(z)$. Therefore
\begin{eqnarray}
C_\ell &\simeq& C_\ell (w^{fid}) + \int{K(\ell,z') \delta w(z')  \ud
z'}\label{eqncellwz}\\
C^{fit}_\ell &\simeq& C_\ell (w^{fid}) + \Delta w \int{K(\ell,z'')
\ud z''}\label{eqncellfit}
\end{eqnarray}
which we will use in the next section.

\subsection{The Weight Function}

The principal goal of this investigation is to quantify the
relationship between the best-fit value of $w$ attained by a survey,
and the underlying function $w(z)$.

This can be obtained by minimising the expectation value of
\begin{equation}
\chi ^2 = \sum_{\ell} \left(
\frac{C^{fit}_\ell-[C_\ell+n(\ell)]}{\sigma(\ell)}\right)^2
,
\end{equation}
where $n(\ell)$ is the experimental noise.  Substituting equations
(\ref{eqncellwz}) and (\ref{eqncellfit}), differentiating with respect
to $w^{\rm fit}$ and averaging over all noise realisations we arrive
at an equation which can be concisely expressed as
\begin{equation} \label{eq:weightint}
w^{fit}=\int{w(z) \Phi(z) dz} .
\end{equation}
Here $w(z)$ is the true, potentially variable, equation of state and
$w^{fit}$ is the average over all noise realisations of the best fit
values for the constant equation of state.  The weight function $\Phi$
is given by
\begin{equation} \label{eq:weight}
\Phi(z) = \frac{\displaystyle \sum_\ell \! \! \int  K(l,z)
K(l,z') /\sigma^2(l) \ud z'}{\displaystyle \sum_\ell \! \! \int
\! \! \! \int K(l,z') K(l,z'') /\sigma^2(l) \ud z' \ud z''} .
\end{equation}

In principle, the weight function may depend on the cosmology, but
within the values of interest ($-1.5<w<-0.5$), the weight function
based on the fiducial model holds to good accuracy, as we shall
soon see.

%
%

\subsection{The Consequences of the $\delta$-function}

In order to evaluate the functional derivative $K(\ell,z)$, we
adopt an equation of state $w(z)=w^{\rm fid}+\epsilon \delta(z'-z)$,
as shown in equation(6), thereby allowing us to rearrange
equations (8,9). Here we consider the impact this perturbation in
$w$ has on the observed shear.

\begin{figure}
\includegraphics[width=3.5in]{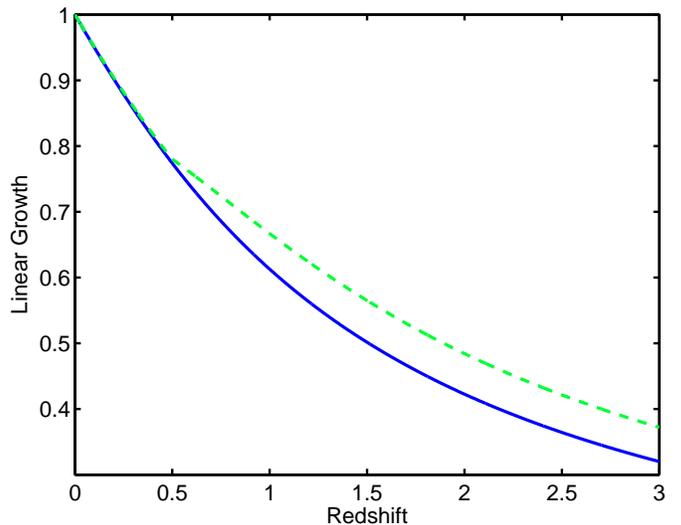}
\caption{\label{fig:Growth} The solid line represents the linear
  growth of structure, normalised at redshift zero. The dotted line
  illustrates the influence of a $\delta$-function in $w$ at $z=0.5$
  ($\epsilon=1$).}
\end{figure}

The dark energy density is given by $\rho_{\rm DE}(z,z')= \rho_{\rm DE}(z=0) f_{\rm DE}(z,z')$ 
where we have defined
\begin{equation}
f_{\rm DE}(z,z') = \left\{ \begin{array}{ll}
 \displaystyle  a^{-3(1+w^{fid})} & z<z'\\
\displaystyle a^{-3(1+w^{fid})}e^{3
\epsilon/(1+z')} & z>z'
\end{array}\right.
\end{equation}
This contributes to a deviation of the growth factor, defined by
$D(z)=\delta(z)/\delta(z=0)$, as illustrated in Fig.
~\ref{fig:Growth}. No modification to the cosmology may occur at
redshifts below $z'$, since we have chosen to constrain the parameters
at redshift zero. At higher redshifts, the enhanced dark energy in
(13) leads to a suppressed rate of growth. This implies a
\emph{greater} prevalence of structure in the past, and we therefore
anticipate a heightened shear signal.

\begin{figure}
\includegraphics[width=3.5in]{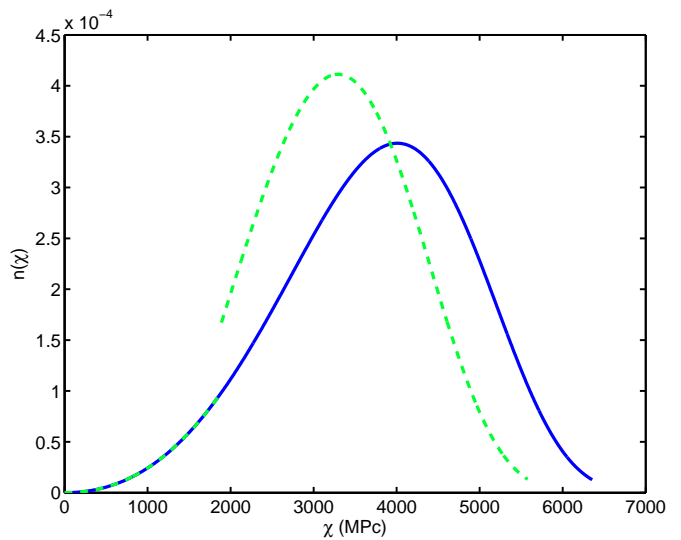}
\caption{\label{fig:Distance} The number of galaxies as a function of 
  coordinate distance, with (dashed) and without (solid) a delta
  function in $w(z)$ perturbing the standard $w=-1$ cosmology ($\epsilon=1$).}
\end{figure}

Conversely, the geometric factors described below suppress the shear signal.
The coordinate distance $\chi(z)$ is now given by:

\begin{equation}
\chi(z,z') =
\frac{c}{H_0} \int_0^{z}\frac{\ud z''}
{  \sqrt{\Omega_m(1+z'')^3+\Omega_{\rm DE}f_{\rm DE}(z',z'')}} .
\end{equation}

In the context of equation (\ref{eq:shear}), the most significant
change made by this modified distance-redshift relation is a shift in
the galaxy probability distribution $n(\chi)$. A cosmic shear survey
will typically use photometric redshift information to evaluate
$n(z)$, whereas the shear signal itself is sensitive to $n(\chi)$,
where $n(\chi)\ud\chi = n(z)\ud z$.  Fig. \ref{fig:Distance}
demonstrates how a $\delta$-function shifts $n(\chi)$ to be closer,
for a given $n(z)$.  Subsequently, the shear is reduced, since it
accumulates over a shorter distance, and the lensing efficiency is
lowered.

In summary, with a more positive w(z), the lensing signal is supported
by the strengthened presence of structure, but reduced by the effects
on the lens geometry.

\begin{figure}
\includegraphics[width=3.5in]{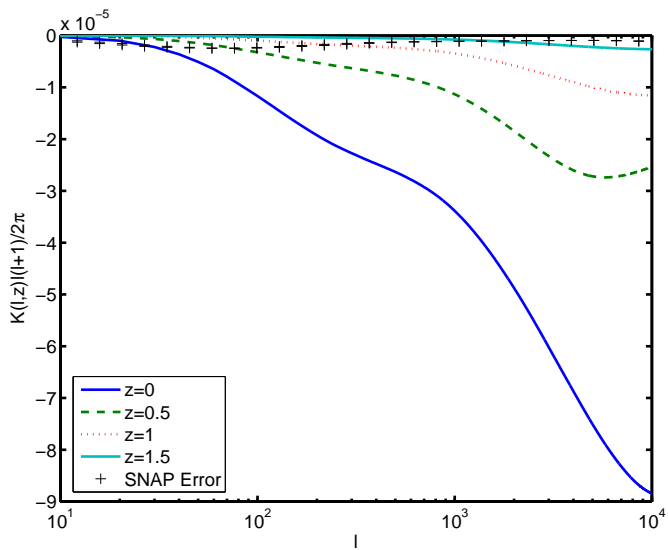}
\caption{\label{fig:Kernel} The kernel at z=0,0.5,1, and 1.5,
  along with the anticipated band-averaged errors for the SNAP wide
  survey. This illustrates the net decrement in the observed shear
  signal due to a positive perturbation in $w(z)$. By considering the
  nature of equation (\ref{eq:weight}), one can begin to visualise the
  weight function taking shape.}
\end{figure}

\begin{figure}
\includegraphics[width=3.5in]{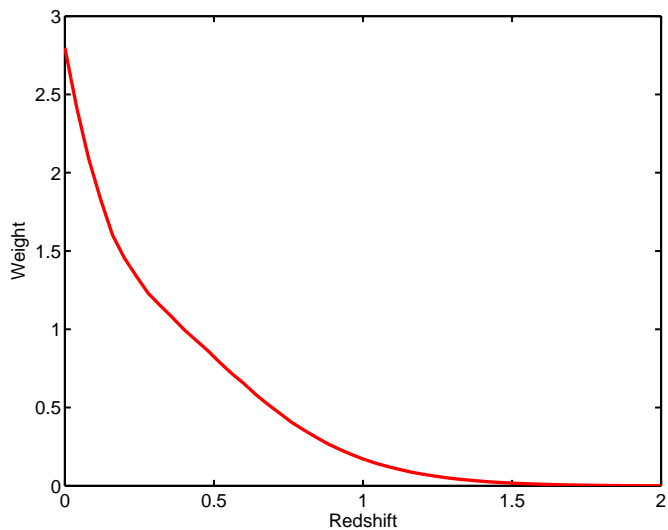}
\caption{\label{fig:SNAP} The weight function corresponding to the
fiducial cosmic shear survey (SNAP wide)
using a single redshift bin.}
\end{figure}

\section{Analysis of Results}

In this section, our aim is to improve our understanding of how cosmic
shear detects dark energy by utilising the weight function approach
described in \S IIC.

\subsection{The Kernel}

Selected redshift slices from $K(\ell,z)$, the functional derivative
of the shear power spectrum, are shown in Fig.~\ref{fig:Kernel}. Here
we follow convention by multiplying the shear $C_\ell$ by a factor of
$\ell(\ell+1)/2\pi$. Therefore, since we have amplified the higher
multipoles, it is not surprising to see that this is where the
greatest change occurs. The behaviour of these multipoles, which
correspond to lensing by the non-linear regime of the matter power
spectrum, will determine the form of the weight function due to their
large signal-to-noise ratio.

The fact that the kernel is negative suggests that there has been a
net decrease in the observed shear, and thus the modification of the
comoving angular diameter distance has prevailed over the matter power
spectrum, which is strengthened by the decreased growth rate.

\subsection{The Weight Function}
\label{sec:weightfn}

Let us now consider the nature of the weight function for the fiducial
cosmic shear survey (SNAP) described in \S\ref{sec:SNAPpars}, as shown
in Fig.~\ref{fig:SNAP}.

The reason for the peak at redshift zero is twofold. Clearly, this is
the point at which dark energy is most prevalent - the sensitivity of
cosmic shear is primarily governed by the decay of $\Omega_{\rm DE}$ -
but another contributing factor is that we have fixed our parameters
at redshift zero, and so by introducing a perturbation in $w$ at a
lower redshift, the cosmology will be modified across a wider redshift
range. We therefore anticipate that using the data to constrain
parameters such as $\Omega_m$, would strengthen the weight function at
higher redshifts.

Another feature of this plot is the fairly abrupt change of gradient
at around $z\sim0.25$.  To explain this we will first have to consider
the separate contributions from structure and geometry.

\begin{figure}
\includegraphics[width=3.5in]{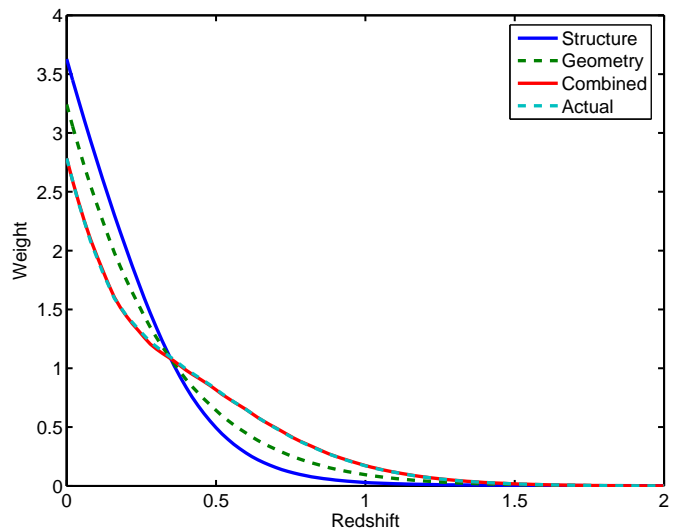}
\caption{\label{fig:strvgro} These weight functions arise when we
  consider modifying the geometric terms and structure growth
  independently. We then use a weighted combination of the two, as
  descibed in the text, to see how the form of the complete weight
  function emerges. The actual weight function is also provided to
  illustrate the accuracy of this approach.}
\end{figure}

We calculated a weight function which corresponds to the influence
felt by cosmic shear via changes in the growth of structure. This was
evaluated by only inserting the delta function into the equation of
state experienced by the matter power spectrum.  This approach was
reapplied for the comoving angular diameter distance.  In Fig.
\ref{fig:strvgro}, the weight functions for these two effects are
plotted.

The weighting functions corresponding to geometry and structure growth
appear quite similar and are both relatively
straightforward, with no sudden changes in gradient.

The structure growth weight function is slightly more strongly peaked
at lower redshift and this is important in understanding the full
cosmic shear weight function.  We interpret this focus on lower
redshift as being due to two factors. Firstly the lensing efficiency
(Equation~\ref{eq:lenseff}) peaks at a redshift \emph{intermediate}
between us and the background galaxies. Secondly due to growth of
structure the conglomeration of matter is more prevalent at low
redshift.  This becomes particularly apparent when we recall that the
matter power spectrum is proportional to the square of the growth
factor.

Additionally, the nature of the nonlinear correction is such that the
higher the power spectrum, the greater the factor of amplification.
For example, at $k\sim1Mpc$ and $z=0$, a 5\% increase in linear growth
can typically lead to a 15\% increase in power.  Therefore $w(z)$
imparts the greatest influence on low-redshift nonlinear structure.

Since the structure growth and geometric influences work in opposite
directions on the cosmic shear power spectrum then the total weight
function is expected to be roughly some \emph{difference} between
these two weighting functions.  The line marked ``Combined" in Fig.
\ref{fig:strvgro} is generated by subtracting the structural weight
function from a version of the geometric one, scaled-up in accordance
with its greater influence. This factor is given by the ratio of
$\partial C_\ell/\partial w_{DA}$ to $\partial C_\ell/\partial w_G$.
From \S\ref{sec:geoorgrowth} this is found to be 1.84. Combinining
weight functions in this manner is valid when the dominant multipoles
(those with the highest signal-to-noise) exhibit the same response to
the equation of state. We find this to be a remarkably accurate
reproduction of the SNAP weight function, thereby verifying our
earlier analysis.

\begin{figure}
\includegraphics[width=3.5in]{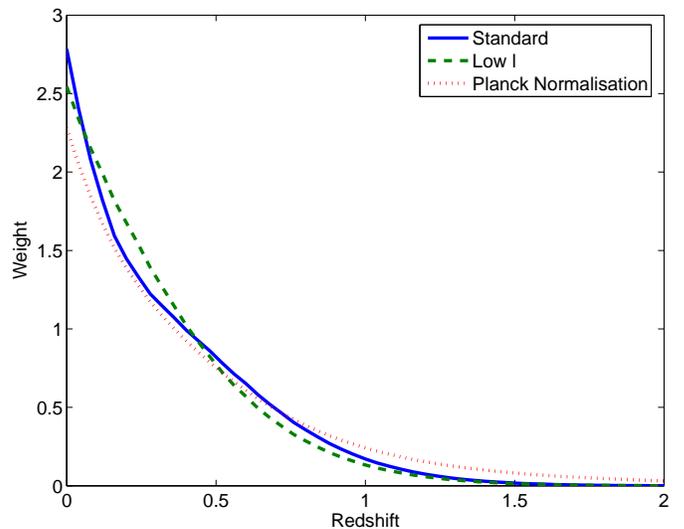}
\caption{\label{fig:Lowl} Here we experiment with alternative constraints. The
  solid line is the standard SNAP weight function from Figure
  \ref{fig:SNAP}.  The dashed line represents the effect of enforcing
  a more conservative upper limit of $\ell<10^3$. The dotted line
  illustrates the effect of constraining the amplitude of the matter
  power spectrum at $z \sim 1000$ instead of $z=0$, as may be achieved
  with anticipated data from Planck}
\end{figure}

Finally, we investigate the effects of modifying our constraints.
Fig. \ref{fig:Lowl} illustrates the response of the weight function to
a reduction in the upper multipole limit ($\ell<1,000$), as would be
necessary in the case of severe baryon contamination. Now we are no
longer sensitive to the non-linear regime, and have thus become
dominated by the geometric signal.

In order to assess the consequences of normalising the growth factor
at high redshift, we utilise the CMB instead of $\sigma_8$ to fix the
amplitude of the matter power spectrum. As illustrated by the dotted
line in \ref{fig:Lowl} the weight function is pushed to slightly
higher redshifts. Note that this rise is attenuated by the loss of the
aforementioned cancellation effect between stucture and geometry. By
constraining the growth factor at high redshift, we have reversed the
effect of the $\delta$ function observed in Fig. \ref{fig:Growth}.
Therefore a slower rate of structure growth will now reduce the
prevalence of structure.

\subsection{Accuracy}

The weight function is only expected to be completely accurate when
the true equation of state $w(z)$ is very close to the fiducial model.
To test the accuracy of the weight function we compare the best fit
value of $w$ (attained by minimisation of $\chi^2$) with that from
integration over the weight function (\ref{eq:weightint}). The true
parameterisation is taken to be of the form $w(z)=w_0+w_a(1-a)$, and
we consider the range $-1.5<w_0<-0.5$ and $-1<w_a<1$.

\begin{figure}
\includegraphics[width=3.5in]{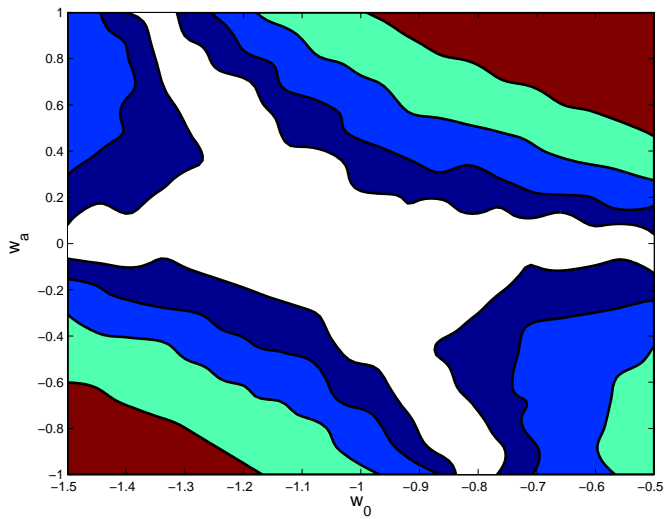}
\caption{\label{fig:Testing} The contours above correspond to
discrepancies of 0.002, 0.005, 0.01 and 0.02 between the the true
value of $w^{\rm fit}$, and that attained by integration over the
weight function.}
\end{figure}

The resulting errors displayed in Fig. \ref{fig:Testing} illustrate
that the weight function approach is sufficiently accurate to apply to
any shear survey currently under consideration.  The perfect accuracy
attained when $w_a=0$ should be regarded as an artefact of the
calculation, since the form of the weight function becomes redundant.
However it is notable that even within the extreme regions of the
plot, much of which lies beyond current observational constraints, the
weight function remains an excellent predictor of the fitted value.

If $w$ is strongly
negative then we would expect the weight function to fall off more
rapidly, in accordance with the decline of $\Omega_{\rm DE}$. While
this effect can be seen in the plot, where the diagonal on which
$w^{\rm fit}=-1$ provides higher accuracy, the deviation elsewhere is
not of great significance.

Nonetheless, it may be worthwhile to reevaluate the weight function in
future, using the observed $w^{\rm fit}$ as the fiducial model.

\section{Tomography}

By dividing the pool of source galaxies into two redshift bins, we can
produce shear power spectra from the two auto-correlations, and a
cross-correlation. This provides us with invaluable information
regarding the evolution of the shear signal, and can considerably
improve constraints on various cosmological parameters. However, the
gain from further subdivision of bins is limited due to their
increasingly correlated shear signals \cite{1999ApJ...522L..21H}.

\begin{figure}
\includegraphics[width=3.5in]{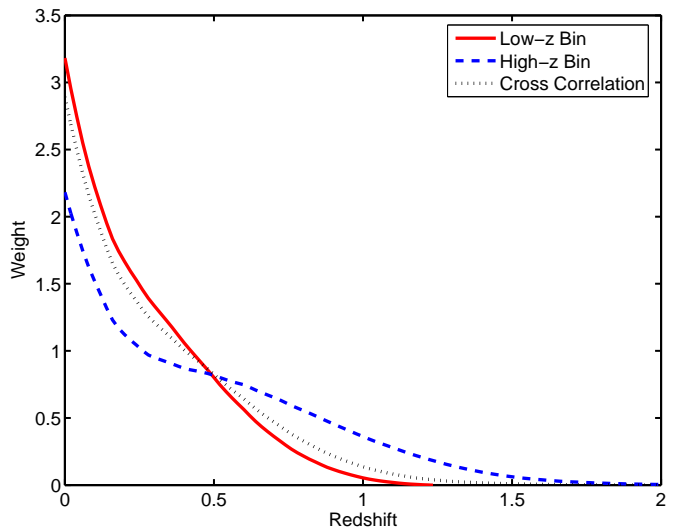}
\caption{\label{fig:TomoWeight} 
  The weight functions for the two redshift bins, and their
  cross-correlation.}
\end{figure}

To evaluate the cross-correlation, the necessary modification to
equation (2) is achieved by replacing $g(\chi)^2$ with $g_A(\chi)
g_B(\chi)$ where A and B are the two redshift bins, split at
approximately $z=1.43$. Following Refregier
\cite{2004AJ....127.3102R}, we produce our binned galaxy
distributions, $n(\chi)$, by modifying the constants and applying a
smear factor to equation (5). Fig.~\ref{fig:TomoWeight} compares the
weight functions of the three power spectra.

In the high-z bin, we see further enhancement of the gradient change
seen in Fig.~\ref{fig:SNAP}, as the perturbed structure plays an
increasingly important role.

\section{Distance or Growth?}
\label{sec:geoorgrowth}

In \S 2 we noted how cosmic shear can reveal dark energy via two
mechanisms. A change in our expansion history can alter both the
distance-redshift relation, and structure growth. In terms of optics,
this is analogous to a modification of lens geometry and focal length
respectively.

Hereafter, we treat these influences separately, by introducing the
effective equations of state $w_G$ and $w_{DA}$. These refer to the
equations of state experienced by growth of structure and comoving
angular diameter distance respectively. Can cosmic shear differentiate
between these quantities?

\begin{figure}
\includegraphics[width=3.5in]{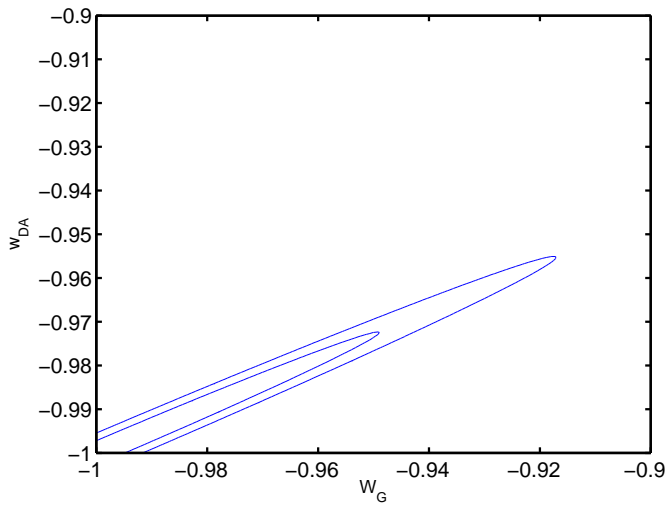}
\caption{\label{fig:Contour} 
  Contours on the dark energy parameter split into its contributions
  to the growth factor and the angular diameter distance (all other
  cosmological parameters are kept fixed). The degeneracy direction
  shows that the two effects partially cancel each other, and that the
  equation of state is slightly better constrained via the angular
  diameter distance.}
\end{figure}

We utilise the standard Fisher formalism to produce the contour
plot for the SNAP-like shear survey shown in Fig.~\ref{fig:Contour}
\begin{equation}
F_{\alpha \beta} = \sum_l \frac{\displaystyle\partial
C_l}{\displaystyle\partial p_{\alpha}}\frac{\displaystyle\partial
C_l}{\displaystyle\partial p_{\beta}}/\sigma^2(l).
\end{equation}

While we can derive little from the size of the contours, since we
have not marginalised over any other parameters, the correlated errors
reinforce several points.

The error in $w_G$ is nearly twice that of $w_{DA}$, while the
orientation of the ellipse is such that their values clearly have
opposing effects on observations. This verifies the findings of
\S~\ref{sec:weightfn} using an independent approach, and is in
agreement with Abazajian \& Dodelson \cite{2003PhRvL..91d1301A}.

\section{Comparison of Lensing Surveys}

We now vary the survey parameters in order to see how the weight
function is affected.  First, we consider a survey originating from
the highest possible redshift, lensing of the Cosmic Microwave
Background (CMB).  At the other end of the spectrum, the CFHTLS will
sample a relatively low redshift distribution of galaxies.

\subsection{The Cosmic Microwave Background}

Lensing of the CMB can in principle be estimated from observations of
higher order moments of the temperature maps, as well as from
polarization observations (see e.g. Hu \& Okamoto
\cite{2002ApJ...574..566H} for more information).

To evaluate the cosmic shear experienced by the CMB, we create an
infinitesimal redshift bin at $z\sim1100$. This simplifies the lensing
efficiency to
\begin{equation}
g(\chi) = 2 \frac{\displaystyle \chi (\chi_{\rm rec} -
\chi)}{\displaystyle \chi_{\rm rec}},
\end{equation}
where $\chi_{\rm rec}$ is the distance to $z=1100$.  Note that the
relevant observable is now the deflection power spectrum, although
this only differs from the standard convergence power spectrum by a
factor of 4/$\ell(\ell+1)$.  The anticipated errors of Planck are
used, as shown in Figure 3 of Hu \cite{2002PhRvD..65b3003H}.

The resulting weight function is shown by the dotted line in Fig.
\ref{fig:lensing}. It is somewhat counter-intuitive to see that the
CMB weight function is less sensitive than SNAP within the
high-redshift regime. Once again, the answer lies in the combination
of geometric and structural factors. Note that for the CMB, $\chi_{\rm
  rec}$ is so large that $w$ has a minimal impact on the lensing
efficiency at low $\chi$, since $g(\chi)\simeq 2 \chi$ and the
distances cancel on substituting this into Equation~\ref{eq:shear}.
Thus, when we perturb $w$, the change in the growth of structure is
the dominant factor, and so we lose the partial cancellation effect
which had pushed the SNAP weight function to higher redshifts.

\subsection{Ground-Based Lensing}

Here we assess two future projects, whose anticipated survey
parameters are outlined below.

\begin{figure}
\includegraphics[width=3.5in]{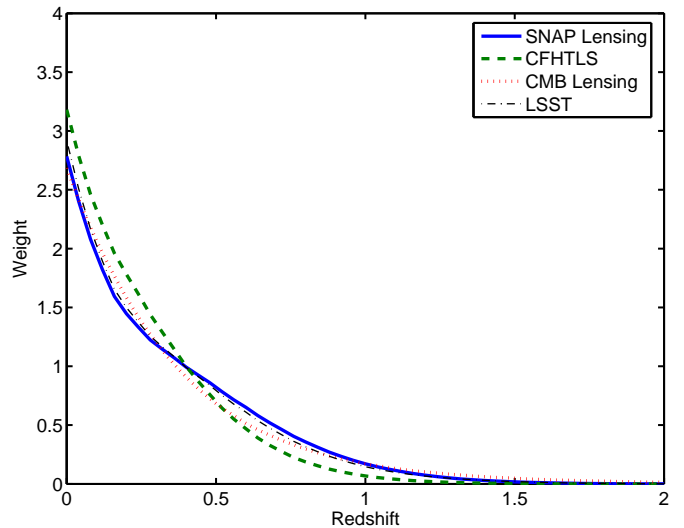}
\caption{\label{fig:lensing} A comparison of lensing surveys. 
  Despite the wide variety of parameters, as outlined in the text
  below, the differences in the weight functions are minor.  The
  dot-dash line corresponding to the LSST is barely distinguishable
  from the SNAP one.  }
\end{figure}

For the CFHTLS, the number denisty of galaxies is reduced to $n_g=26$ ${\rm
  arcmin}^{-2}$. The galaxy redshift distribution, $n(z)$, is taken
from the VIRMOS estimation by van Waerbeke et al.
\cite{2002A&A...393..369V}. In the context of equation \ref{eq:nofz}
we now have $\alpha=2$, $\beta=1.2$, and $z_0=0.44$.  The rms shear
$\sigma_\gamma=0.31$, and the area of the survey is 170 square degrees. The
resulting weight function is shown by the dashed line in Fig.
\ref{fig:lensing}. The sensitivity at high redshift is limited by the
depth of the survey.

For the LSST, the number denisty of galaxies $n_g=65$ ${\rm
  arcmin}^{-2}$, and the rms shear $\sigma_\gamma=0.16$ .  The
redshift distribution is adopted from Song \& Knox
\cite{2004PhRvD..70f3510S}. The resulting weight function is the
dot-dash line in Fig.  \ref{fig:lensing}. The close resemblance to the
SNAP plot is not surprising given their similar redshift
distributions. And while the survey is much wider (30,000 square
degrees), it should be noted that the weight function is in fact
\emph{independent} of the sky coverage, since this just enters as a
constant multiple of the errors.

\section{Comparing Alternate Probes of Dark Energy}

We turn our attention to Fig.~\ref{fig:weightfuns} which shows the
response of different methods to the equation of state.

\subsection{Benchmark}

The `benchmark' is the weight attributed to a purely hypothetical
survey which equally examines dark energy within all redshift bins.
That is to say, it represents a neutral (albeit somewhat unphysical)
observation whereby the value of $w$ for each quanta of dark energy is
measured.

In the context of equation (\ref{eq:weight}), we have no redshift
dependency so $K(l,z)={\rm const.}$, and $\sigma^2(z) \propto
(1+z)^{3}$ due to the amount of dark energy present.

This leads us to a weight function which is a (normalised) plot of volume 
versus redshift.

\begin{equation}
\Phi_{bench}(z) = \frac{\displaystyle 2}{\displaystyle (1+z)^3}
\end{equation}

This is taken as a reference point, from which we can identify the
redshift bias of more complex methods.  We see that a cosmic shear
survey with the SNAP parameters comes quite close to this benchmark.

\subsection{Type 1A Supernovae}

\begin{figure}
\includegraphics[width=3.5in]{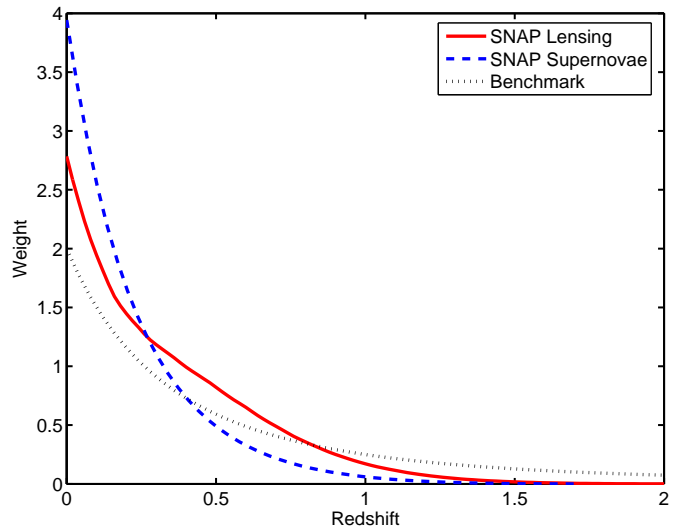}
\caption{\label{fig:weightfuns} The weight functions for lensing
  of the cosmic shear (solid), supernovae evenly distributed from
  $z=0.1$ to $z=1.7$ (dashed), and benchmark (dotted). Cosmic shear
  probes the equation of state at higher redshifts than the supernovae
  (for SNAP-like surveys), as explained in the text.}
\end{figure}

The dashed line in Fig.~\ref{fig:weightfuns} is calculated for the
proposed SNAP supernovae study, where supernovae are assumed to be
uniformly distributed between redshifts 0.1 and 1.7, and errors are
taken to be inversely proportional to the coordinate distance.

One might intuitively expect supernovae to probe higher redshift
regions than cosmic shear, since lensing mostly occurs approximately
mid-way between source and observer. However, Section IID has already
outlined why the combination of structural and geometric factors have
pushed the sensitivity to higher $z$.  Moreover, the poor accuracy of
high-$z$ supernovae has the opposite effect. The resulting weight
function is shown in Fig.  \ref{fig:weightfuns} and was also
calculated in Saini et al \cite{2003MNRAS.343..533D}.  It should be
noted that even a high-redshift supernovae survey actually measures
$w(z)$ at a redshift lower than a shallow cosmic shear survey.

Let us now consider a possible consequence of surveys which have
significantly different weight functions. Disparity in the
constant-$w$ fits may not necessarily indicate inconsistent data. For
instance, if we adopt a varying equation of state of the form
$w=w_0+w_a(1-a)$ then in order for $w^{fit}_{Shear}$ to differ from
$w^{fit}_{SNAe}$ by 5\% we would require $|w_a|>0.7|w_0|$.

\section{Discussion}

We have revealed the manner in which cosmic shear probes the dark
energy equation of state. The survey specifications of SNAP, due for
launch in 2014, have been taken as an example. However, it should be
noted that in general, weight functions are invariant to the absolute
errors of the observation, and as such the survey-dependency of the
weight function is small.

The important point is that interpreting the equation of state via
weight functions circumvents the problem of which parameters to use
for the evolution of the dark energy.  A parameterization can be
decided on, but if the true equation of state does not have the
functional form assumed, the meaning of the coefficients can still be
interpreted clearly in terms of the true variation of the equation of
state.

The weight function for cosmic shear, which describes the redshifts to
which we are sensitive, is found to differ significantly from that of
a supernovae survey. Therefore a discrepancy in the values of $w$ may
be indicative of an evolving equation of state.

Remarkably, the competing factors of structure and geometry ensure
that cosmic shear will be more sensitive to the high-redshift values
of $w(z)$ than lensing of the CMB. Since the primary goal is to detect
a deviation from $w=-1$, and many theories suggest that this would be
most prominent at high redshift, then these high-z sensitivities are
particularly desirable.

The way in which $w$ modifies the nonlinear correction to the matter
power spectrum remains an issue, and we await the outcome of large
N-body simulations of quintessence cosmologies in order to progress
further.

The principal component analysis of Huterer \cite{2003PhRvL..90c1301H}
also investigates the sensitivity of supernova data to the equation of
state as a function of redshift, focussing on finding the modes that
can be well constrained by the data.  The extension of this work to
compare supernova and weak lensing data by Knox et al.
\cite{2004AAS...20510814K} supports our findings.

This work is complementary to that of Hu (2004)
\cite{astro-ph/0407158} which analyses the change of various measures
(e.g. angular diameter distance as a function of redshift) on changing
w from a cosmological constant to an alternative model, at the same
time modifying the amount of dark energy $\Omega_{\rm DE}$ to keep the
angular diameter distance to the CMB constant. We have fixed the
cosmological parameters at redshift zero for the current analysis and
an investigation into the multitude of alternative priors is beyond
the scope of this work. Our approach differs from the investigation of
Hu in that it is not dependent on specific modifications to the dark
energy equation of state, and investigates cosmic shear specifically.

There are a number of routes down which this work could progress.
Specifically, the fitting of extra parameters such as $w_0$ and $w_1$
(for $w(z) = w_0 + w_1 z$) which has been done for the supernova
weight function in \cite{2003MNRAS.343..533D} or a reconstruction of
$w(z)$ in redshift bins. This could even be extended to other
cosmological parameters such as $\Omega_m$. Maor et al.
\cite{2002PhRvD..65l3003M} have highlighted the problems associated
with determining $w$ when $\Omega_m$ is not tightly constrained. In
addition, it would be instructive to compare the standard cosmic shear
weight function with that of the Jain \& Taylor
\cite{2003PhRvL..91n1302J} approach.

The illustrative nature of weight functions provides an intuitive
insight into the meaning of experimental results. Furthermore, we
believe this approach will prove useful in optimising future
surveillance strategies. For instance, it may become desirable to
probe a particular redshift regime of $w(z)$, at which point the
weight functions of various surveys should be considered. This is most
readily achieved by careful selection of the source distribution n(z).
However, in doing so we must ensure the overall accuracy of the
measurement is not compromised.

\begin{acknowledgments}
  We thank T. Saini, L. King, D. Bacon, M. Rees, G. Bernstein, L.
  Knox, F. Abdalla and A. Refregier for helpful comments. FRGS
  acknowledges the support of Trinity College. SLB thanks the Royal
  Society for support in the form of a University Research Fellowship.
\end{acknowledgments}
\bibliography{bibs}

\begin{thebibliography}{30}
\expandafter\ifx\csname natexlab\endcsname\relax\def\natexlab#1{#1}\fi
\expandafter\ifx\csname bibnamefont\endcsname\relax
  \def\bibnamefont#1{#1}\fi
\expandafter\ifx\csname bibfnamefont\endcsname\relax
  \def\bibfnamefont#1{#1}\fi
\expandafter\ifx\csname citenamefont\endcsname\relax
  \def\citenamefont#1{#1}\fi
\expandafter\ifx\csname url\endcsname\relax
  \def\url#1{\texttt{#1}}\fi
\expandafter\ifx\csname urlprefix\endcsname\relax\def\urlprefix{URL }\fi
\providecommand{\bibinfo}[2]{#2}
\providecommand{\eprint}[2][]{\url{#2}}

\bibitem[{\citenamefont{{Deep Saini} et~al.}(2003)\citenamefont{{Deep Saini},
  {Padmanabhan}, and {Bridle}}}]{2003MNRAS.343..533D}
\bibinfo{author}{\bibfnamefont{T.}~\bibnamefont{{Deep Saini}}},
  \bibinfo{author}{\bibfnamefont{T.}~\bibnamefont{{Padmanabhan}}},
  \bibnamefont{and} \bibinfo{author}{\bibfnamefont{S.}~\bibnamefont{{Bridle}}},
  \bibinfo{journal}{MNRAS} \textbf{\bibinfo{volume}{343}}, \bibinfo{pages}{533}
  (\bibinfo{year}{2003}).

\bibitem[{\citenamefont{{Perlmutter} et~al.}(1999)\citenamefont{{Perlmutter},
  {Aldering}, {Goldhaber}, {Knop}, {Nugent}, {Castro}, {Deustua}, {Fabbro},
  {Goobar}, {Groom} et~al.}}]{1999ApJ...517..565P}
\bibinfo{author}{\bibfnamefont{S.}~\bibnamefont{{Perlmutter}}},
  \bibinfo{author}{\bibfnamefont{G.}~\bibnamefont{{Aldering}}},
  \bibinfo{author}{\bibfnamefont{G.}~\bibnamefont{{Goldhaber}}},
  \bibinfo{author}{\bibfnamefont{R.~A.} \bibnamefont{{Knop}}},
  \bibinfo{author}{\bibfnamefont{P.}~\bibnamefont{{Nugent}}},
  \bibinfo{author}{\bibfnamefont{P.~G.} \bibnamefont{{Castro}}},
  \bibinfo{author}{\bibfnamefont{S.}~\bibnamefont{{Deustua}}},
  \bibinfo{author}{\bibfnamefont{S.}~\bibnamefont{{Fabbro}}},
  \bibinfo{author}{\bibfnamefont{A.}~\bibnamefont{{Goobar}}},
  \bibinfo{author}{\bibfnamefont{D.~E.} \bibnamefont{{Groom}}},
  \bibnamefont{et~al.}, \bibinfo{journal}{\apj} \textbf{\bibinfo{volume}{517}},
  \bibinfo{pages}{565} (\bibinfo{year}{1999}).

\bibitem[{\citenamefont{{Allen} et~al.}(2004)\citenamefont{{Allen}, {Schmidt},
  {Ebeling}, {Fabian}, and {van Speybroeck}}}]{2004MNRAS.tmp..258A}
\bibinfo{author}{\bibfnamefont{S.~W.} \bibnamefont{{Allen}}},
  \bibinfo{author}{\bibfnamefont{R.~W.} \bibnamefont{{Schmidt}}},
  \bibinfo{author}{\bibfnamefont{H.}~\bibnamefont{{Ebeling}}},
  \bibinfo{author}{\bibfnamefont{A.~C.} \bibnamefont{{Fabian}}},
  \bibnamefont{and} \bibinfo{author}{\bibfnamefont{L.}~\bibnamefont{{van
  Speybroeck}}}, \bibinfo{journal}{MNRAS} pp. \bibinfo{pages}{258--+}
  (\bibinfo{year}{2004}).

\bibitem[{\citenamefont{{Tegmark} et~al.}(2004)\citenamefont{{Tegmark},
  {Strauss}, {Blanton}, {Abazajian}, {Dodelson}, {Sandvik}, {Wang}, {Weinberg},
  {Zehavi}, {Bahcall} et~al.}}]{2004PhRvD..69j3501T}
\bibinfo{author}{\bibfnamefont{M.}~\bibnamefont{{Tegmark}}},
  \bibinfo{author}{\bibfnamefont{M.~A.} \bibnamefont{{Strauss}}},
  \bibinfo{author}{\bibfnamefont{M.~R.} \bibnamefont{{Blanton}}},
  \bibinfo{author}{\bibfnamefont{K.}~\bibnamefont{{Abazajian}}},
  \bibinfo{author}{\bibfnamefont{S.}~\bibnamefont{{Dodelson}}},
  \bibinfo{author}{\bibfnamefont{H.}~\bibnamefont{{Sandvik}}},
  \bibinfo{author}{\bibfnamefont{X.}~\bibnamefont{{Wang}}},
  \bibinfo{author}{\bibfnamefont{D.~H.} \bibnamefont{{Weinberg}}},
  \bibinfo{author}{\bibfnamefont{I.}~\bibnamefont{{Zehavi}}},
  \bibinfo{author}{\bibfnamefont{N.~A.} \bibnamefont{{Bahcall}}},
  \bibnamefont{et~al.}, \bibinfo{journal}{\prd} \textbf{\bibinfo{volume}{69}},
  \bibinfo{pages}{103501} (\bibinfo{year}{2004}).

\bibitem[{\citenamefont{{Spergel} et~al.}(2003)\citenamefont{{Spergel},
  {Verde}, {Peiris}, {Komatsu}, {Nolta}, {Bennett}, {Halpern}, {Hinshaw},
  {Jarosik}, {Kogut} et~al.}}]{2003ApJS..148..175S}
\bibinfo{author}{\bibfnamefont{D.~N.} \bibnamefont{{Spergel}}},
  \bibinfo{author}{\bibfnamefont{L.}~\bibnamefont{{Verde}}},
  \bibinfo{author}{\bibfnamefont{H.~V.} \bibnamefont{{Peiris}}},
  \bibinfo{author}{\bibfnamefont{E.}~\bibnamefont{{Komatsu}}},
  \bibinfo{author}{\bibfnamefont{M.~R.} \bibnamefont{{Nolta}}},
  \bibinfo{author}{\bibfnamefont{C.~L.} \bibnamefont{{Bennett}}},
  \bibinfo{author}{\bibfnamefont{M.}~\bibnamefont{{Halpern}}},
  \bibinfo{author}{\bibfnamefont{G.}~\bibnamefont{{Hinshaw}}},
  \bibinfo{author}{\bibfnamefont{N.}~\bibnamefont{{Jarosik}}},
  \bibinfo{author}{\bibfnamefont{A.}~\bibnamefont{{Kogut}}},
  \bibnamefont{et~al.}, \bibinfo{journal}{\apjs}
  \textbf{\bibinfo{volume}{148}}, \bibinfo{pages}{175} (\bibinfo{year}{2003}).

\bibitem[{\citenamefont{{Van Waerbeke} et~al.}(2000)\citenamefont{{Van
  Waerbeke}, {Mellier}, {Erben}, {Cuillandre}, {Bernardeau}, {Maoli}, {Bertin},
  {Mc Cracken}, {Le F{\` e}vre}, {Fort} et~al.}}]{2000A&A...358...30V}
\bibinfo{author}{\bibfnamefont{L.}~\bibnamefont{{Van Waerbeke}}},
  \bibinfo{author}{\bibfnamefont{Y.}~\bibnamefont{{Mellier}}},
  \bibinfo{author}{\bibfnamefont{T.}~\bibnamefont{{Erben}}},
  \bibinfo{author}{\bibfnamefont{J.~C.} \bibnamefont{{Cuillandre}}},
  \bibinfo{author}{\bibfnamefont{F.}~\bibnamefont{{Bernardeau}}},
  \bibinfo{author}{\bibfnamefont{R.}~\bibnamefont{{Maoli}}},
  \bibinfo{author}{\bibfnamefont{E.}~\bibnamefont{{Bertin}}},
  \bibinfo{author}{\bibfnamefont{H.~J.} \bibnamefont{{Mc Cracken}}},
  \bibinfo{author}{\bibfnamefont{O.}~\bibnamefont{{Le F{\` e}vre}}},
  \bibinfo{author}{\bibfnamefont{B.}~\bibnamefont{{Fort}}},
  \bibnamefont{et~al.}, \bibinfo{journal}{\aap} \textbf{\bibinfo{volume}{358}},
  \bibinfo{pages}{30} (\bibinfo{year}{2000}).

\bibitem[{\citenamefont{{Bacon} et~al.}(2000)\citenamefont{{Bacon},
  {Refregier}, and {Ellis}}}]{2000MNRAS.318..625B}
\bibinfo{author}{\bibfnamefont{D.~J.} \bibnamefont{{Bacon}}},
  \bibinfo{author}{\bibfnamefont{A.~R.} \bibnamefont{{Refregier}}},
  \bibnamefont{and} \bibinfo{author}{\bibfnamefont{R.~S.}
  \bibnamefont{{Ellis}}}, \bibinfo{journal}{\mnras}
  \textbf{\bibinfo{volume}{318}}, \bibinfo{pages}{625} (\bibinfo{year}{2000}).

\bibitem[{\citenamefont{{Wittman} et~al.}(2000)\citenamefont{{Wittman},
  {Tyson}, {Kirkman}, {Dell'Antonio}, and {Bernstein}}}]{2000Natur.405..143W}
\bibinfo{author}{\bibfnamefont{D.~M.} \bibnamefont{{Wittman}}},
  \bibinfo{author}{\bibfnamefont{J.~A.} \bibnamefont{{Tyson}}},
  \bibinfo{author}{\bibfnamefont{D.}~\bibnamefont{{Kirkman}}},
  \bibinfo{author}{\bibfnamefont{I.}~\bibnamefont{{Dell'Antonio}}},
  \bibnamefont{and}
  \bibinfo{author}{\bibfnamefont{G.}~\bibnamefont{{Bernstein}}},
  \bibinfo{journal}{\nat} \textbf{\bibinfo{volume}{405}}, \bibinfo{pages}{143}
  (\bibinfo{year}{2000}).

\bibitem[{\citenamefont{{Jarvis} et~al.}(2003)\citenamefont{{Jarvis},
  {Bernstein}, {Fischer}, {Smith}, {Jain}, {Tyson}, and
  {Wittman}}}]{2003AJ....125.1014J}
\bibinfo{author}{\bibfnamefont{M.}~\bibnamefont{{Jarvis}}},
  \bibinfo{author}{\bibfnamefont{G.~M.} \bibnamefont{{Bernstein}}},
  \bibinfo{author}{\bibfnamefont{P.}~\bibnamefont{{Fischer}}},
  \bibinfo{author}{\bibfnamefont{D.}~\bibnamefont{{Smith}}},
  \bibinfo{author}{\bibfnamefont{B.}~\bibnamefont{{Jain}}},
  \bibinfo{author}{\bibfnamefont{J.~A.} \bibnamefont{{Tyson}}},
  \bibnamefont{and}
  \bibinfo{author}{\bibfnamefont{D.}~\bibnamefont{{Wittman}}},
  \bibinfo{journal}{\aj} \textbf{\bibinfo{volume}{125}}, \bibinfo{pages}{1014}
  (\bibinfo{year}{2003}).

\bibitem[{\citenamefont{{Kaiser} et~al.}(2000)\citenamefont{{Kaiser}, {Wilson},
  and {Luppino}}}]{astro-ph/0003338}
\bibinfo{author}{\bibfnamefont{N.}~\bibnamefont{{Kaiser}}},
  \bibinfo{author}{\bibfnamefont{G.}~\bibnamefont{{Wilson}}}, \bibnamefont{and}
  \bibinfo{author}{\bibfnamefont{G.}~\bibnamefont{{Luppino}}},
  \bibinfo{journal}{arXiv:astro-ph/0003338}  (\bibinfo{year}{2000}).

\bibitem[{\citenamefont{{Ma} et~al.}(1999)\citenamefont{{Ma}, {Caldwell},
  {Bode}, and {Wang}}}]{1999ApJ...521L...1M}
\bibinfo{author}{\bibfnamefont{C.}~\bibnamefont{{Ma}}},
  \bibinfo{author}{\bibfnamefont{R.~R.} \bibnamefont{{Caldwell}}},
  \bibinfo{author}{\bibfnamefont{P.}~\bibnamefont{{Bode}}}, \bibnamefont{and}
  \bibinfo{author}{\bibfnamefont{L.}~\bibnamefont{{Wang}}},
  \bibinfo{journal}{\apjl} \textbf{\bibinfo{volume}{521}}, \bibinfo{pages}{L1}
  (\bibinfo{year}{1999}).

\bibitem[{\citenamefont{{Peacock} and {Dodds}}(1996)}]{1996MNRAS.280L..19P}
\bibinfo{author}{\bibfnamefont{J.~A.} \bibnamefont{{Peacock}}}
  \bibnamefont{and} \bibinfo{author}{\bibfnamefont{S.~J.}
  \bibnamefont{{Dodds}}}, \bibinfo{journal}{MNRAS}
  \textbf{\bibinfo{volume}{280}}, \bibinfo{pages}{L19} (\bibinfo{year}{1996}).

\bibitem[{\citenamefont{{Bartelmann} and
  {Schneider}}(1999)}]{1999A&A...345...17B}
\bibinfo{author}{\bibfnamefont{M.}~\bibnamefont{{Bartelmann}}}
  \bibnamefont{and}
  \bibinfo{author}{\bibfnamefont{P.}~\bibnamefont{{Schneider}}},
  \bibinfo{journal}{\aap} \textbf{\bibinfo{volume}{345}}, \bibinfo{pages}{17}
  (\bibinfo{year}{1999}).

\bibitem[{\citenamefont{{Mellier}}(1999)}]{1999ARA&A..37..127M}
\bibinfo{author}{\bibfnamefont{Y.}~\bibnamefont{{Mellier}}},
  \bibinfo{journal}{\araa} \textbf{\bibinfo{volume}{37}}, \bibinfo{pages}{127}
  (\bibinfo{year}{1999}).

\bibitem[{\citenamefont{{Kaiser}}(1998)}]{1998ApJ...498...26K}
\bibinfo{author}{\bibfnamefont{N.}~\bibnamefont{{Kaiser}}},
  \bibinfo{journal}{\apj} \textbf{\bibinfo{volume}{498}}, \bibinfo{pages}{26}
  (\bibinfo{year}{1998}).

\bibitem[{\citenamefont{{Refregier} et~al.}(2004)\citenamefont{{Refregier},
  {Massey}, {Rhodes}, {Ellis}, {Albert}, {Bacon}, {Bernstein}, {McKay}, and
  {Perlmutter}}}]{2004AJ....127.3102R}
\bibinfo{author}{\bibfnamefont{A.}~\bibnamefont{{Refregier}}},
  \bibinfo{author}{\bibfnamefont{R.}~\bibnamefont{{Massey}}},
  \bibinfo{author}{\bibfnamefont{J.}~\bibnamefont{{Rhodes}}},
  \bibinfo{author}{\bibfnamefont{R.}~\bibnamefont{{Ellis}}},
  \bibinfo{author}{\bibfnamefont{J.}~\bibnamefont{{Albert}}},
  \bibinfo{author}{\bibfnamefont{D.}~\bibnamefont{{Bacon}}},
  \bibinfo{author}{\bibfnamefont{G.}~\bibnamefont{{Bernstein}}},
  \bibinfo{author}{\bibfnamefont{T.}~\bibnamefont{{McKay}}}, \bibnamefont{and}
  \bibinfo{author}{\bibfnamefont{S.}~\bibnamefont{{Perlmutter}}},
  \bibinfo{journal}{\aj} \textbf{\bibinfo{volume}{127}}, \bibinfo{pages}{3102}
  (\bibinfo{year}{2004}).

\bibitem[{\citenamefont{{Cooray} and {Hu}}(2001)}]{2001ApJ...554...56C}
\bibinfo{author}{\bibfnamefont{A.}~\bibnamefont{{Cooray}}} \bibnamefont{and}
  \bibinfo{author}{\bibfnamefont{W.}~\bibnamefont{{Hu}}},
  \bibinfo{journal}{\apj} \textbf{\bibinfo{volume}{554}}, \bibinfo{pages}{56}
  (\bibinfo{year}{2001}).

\bibitem[{\citenamefont{{White}}(2004)}]{2004APh....22..211W}
\bibinfo{author}{\bibfnamefont{M.}~\bibnamefont{{White}}},
  \bibinfo{journal}{Astroparticle Physics} \textbf{\bibinfo{volume}{22}},
  \bibinfo{pages}{211} (\bibinfo{year}{2004}).

\bibitem[{\citenamefont{{Zhan} and {Knox}}(2004)}]{2004ApJ...616L..75Z}
\bibinfo{author}{\bibfnamefont{H.}~\bibnamefont{{Zhan}}} \bibnamefont{and}
  \bibinfo{author}{\bibfnamefont{L.}~\bibnamefont{{Knox}}},
  \bibinfo{journal}{\apjl} \textbf{\bibinfo{volume}{616}}, \bibinfo{pages}{L75}
  (\bibinfo{year}{2004}).

\bibitem[{\citenamefont{{Hu}}(1999)}]{1999ApJ...522L..21H}
\bibinfo{author}{\bibfnamefont{W.}~\bibnamefont{{Hu}}},
  \bibinfo{journal}{\apjl} \textbf{\bibinfo{volume}{522}}, \bibinfo{pages}{L21}
  (\bibinfo{year}{1999}).

\bibitem[{\citenamefont{{Abazajian} and
  {Dodelson}}(2003)}]{2003PhRvL..91d1301A}
\bibinfo{author}{\bibfnamefont{K.}~\bibnamefont{{Abazajian}}} \bibnamefont{and}
  \bibinfo{author}{\bibfnamefont{S.}~\bibnamefont{{Dodelson}}},
  \bibinfo{journal}{Physical Review Letters} \textbf{\bibinfo{volume}{91}},
  \bibinfo{pages}{041301} (\bibinfo{year}{2003}).

\bibitem[{\citenamefont{{Hu} and {Okamoto}}(2002)}]{2002ApJ...574..566H}
\bibinfo{author}{\bibfnamefont{W.}~\bibnamefont{{Hu}}} \bibnamefont{and}
  \bibinfo{author}{\bibfnamefont{T.}~\bibnamefont{{Okamoto}}},
  \bibinfo{journal}{\apj} \textbf{\bibinfo{volume}{574}}, \bibinfo{pages}{566}
  (\bibinfo{year}{2002}).

\bibitem[{\citenamefont{{Hu}}(2002)}]{2002PhRvD..65b3003H}
\bibinfo{author}{\bibfnamefont{W.}~\bibnamefont{{Hu}}}, \bibinfo{journal}{\prd}
  \textbf{\bibinfo{volume}{65}}, \bibinfo{pages}{023003}
  (\bibinfo{year}{2002}).

\bibitem[{\citenamefont{{Van Waerbeke} et~al.}(2002)\citenamefont{{Van
  Waerbeke}, {Mellier}, {Pell{\' o}}, {Pen}, {McCracken}, and
  {Jain}}}]{2002A&A...393..369V}
\bibinfo{author}{\bibfnamefont{L.}~\bibnamefont{{Van Waerbeke}}},
  \bibinfo{author}{\bibfnamefont{Y.}~\bibnamefont{{Mellier}}},
  \bibinfo{author}{\bibfnamefont{R.}~\bibnamefont{{Pell{\' o}}}},
  \bibinfo{author}{\bibfnamefont{U.-L.} \bibnamefont{{Pen}}},
  \bibinfo{author}{\bibfnamefont{H.~J.} \bibnamefont{{McCracken}}},
  \bibnamefont{and} \bibinfo{author}{\bibfnamefont{B.}~\bibnamefont{{Jain}}},
  \bibinfo{journal}{\aap} \textbf{\bibinfo{volume}{393}}, \bibinfo{pages}{369}
  (\bibinfo{year}{2002}).

\bibitem[{\citenamefont{{Song} and {Knox}}(2004)}]{2004PhRvD..70f3510S}
\bibinfo{author}{\bibfnamefont{Y.}~\bibnamefont{{Song}}} \bibnamefont{and}
  \bibinfo{author}{\bibfnamefont{L.}~\bibnamefont{{Knox}}},
  \bibinfo{journal}{\prd} \textbf{\bibinfo{volume}{70}},
  \bibinfo{pages}{063510} (\bibinfo{year}{2004}).

\bibitem[{\citenamefont{{Huterer} and {Starkman}}(2003)}]{2003PhRvL..90c1301H}
\bibinfo{author}{\bibfnamefont{D.}~\bibnamefont{{Huterer}}} \bibnamefont{and}
  \bibinfo{author}{\bibfnamefont{G.}~\bibnamefont{{Starkman}}},
  \bibinfo{journal}{Physical Review Letters} \textbf{\bibinfo{volume}{90}},
  \bibinfo{pages}{031301} (\bibinfo{year}{2003}).

\bibitem[{\citenamefont{{Knox} et~al.}(2004)\citenamefont{{Knox}, {Albrecht},
  {Song}, {Tyson}, and {Wittman}}}]{2004AAS...20510814K}
\bibinfo{author}{\bibfnamefont{L.}~\bibnamefont{{Knox}}},
  \bibinfo{author}{\bibfnamefont{A.}~\bibnamefont{{Albrecht}}},
  \bibinfo{author}{\bibfnamefont{Y.-S.} \bibnamefont{{Song}}},
  \bibinfo{author}{\bibfnamefont{J.~A.} \bibnamefont{{Tyson}}},
  \bibnamefont{and}
  \bibinfo{author}{\bibfnamefont{D.}~\bibnamefont{{Wittman}}},
  \bibinfo{journal}{American Astronomical Society Meeting Abstracts}
  \textbf{\bibinfo{volume}{205}},  (\bibinfo{year}{2004}).

\bibitem[{\citenamefont{{Hu}}(2004)}]{astro-ph/0407158}
\bibinfo{author}{\bibfnamefont{W.}~\bibnamefont{{Hu}}},
  \bibinfo{journal}{arXiv:astro-ph/0407158}  (\bibinfo{year}{2004}).

\bibitem[{\citenamefont{{Maor} et~al.}(2002)\citenamefont{{Maor}, {Brustein},
  {McMahon}, and {Steinhardt}}}]{2002PhRvD..65l3003M}
\bibinfo{author}{\bibfnamefont{I.}~\bibnamefont{{Maor}}},
  \bibinfo{author}{\bibfnamefont{R.}~\bibnamefont{{Brustein}}},
  \bibinfo{author}{\bibfnamefont{J.}~\bibnamefont{{McMahon}}},
  \bibnamefont{and} \bibinfo{author}{\bibfnamefont{P.~J.}
  \bibnamefont{{Steinhardt}}}, \bibinfo{journal}{\prd}
  \textbf{\bibinfo{volume}{65}}, \bibinfo{pages}{123003}
  (\bibinfo{year}{2002}).

\bibitem[{\citenamefont{{Jain} and {Taylor}}(2003)}]{2003PhRvL..91n1302J}
\bibinfo{author}{\bibfnamefont{B.}~\bibnamefont{{Jain}}} \bibnamefont{and}
  \bibinfo{author}{\bibfnamefont{A.}~\bibnamefont{{Taylor}}},
  \bibinfo{journal}{Physical Review Letters} \textbf{\bibinfo{volume}{91}},
  \bibinfo{pages}{141302} (\bibinfo{year}{2003}).

\end{thebibliography}

\end{document}